\documentstyle[12pt,amsfonts]{article}
\newcommand{\ehat}{ \hat U_{\epsilon} }

\newcommand{\define}{ \stackrel{\triangle}{=} }

\def\be{\begin{equation}}
\def\ee{\end{equation}}
\def\ba{\begin{array}}
\def\ea{\end{array}}

\def\d4{{\rm d}^4}

\begin{document}
\title{\bf Unified Theory of Fundamental Interactions }
\author{{Ning Wu}
\thanks{email address: wuning@mail.ihep.ac.cn}
\\
\\
{\small Institute of High Energy Physics, P.O.Box 918-1,
Beijing 100039, P.R.China}}
\maketitle
\vskip 0.8in

~~

\vskip 0.8in

\begin{abstract}

Based on local gauge invariance, four different kinds of 
fundamental interactions in Nature  are unified in a 
theory which has $SU(3)_c
\otimes SU(2)_L \otimes U(1) \otimes _s Gravitational~ Gauge~ Group$
gauge symmetry. In this approach, gravitational field, like 
electromagnetic field, intermediate gauge field and gluon
field, is represented by gauge potential. Four kinds of
fundamental interactions are formulated in the similar
manner, and therefore can be unified in a direct or
semi-direct product group. The model discussed in 
this paper can be regarded as extension of the 
standard model to gravitational interactions. 
The model discussed in this paper is a renormalizable
quantum model, so it can be used to study quantum effects 
of gravitational interactions.

\end{abstract}

\newpage

\Roman{section}

\section{Introduction}

It is known that there are four kinds of fundamental interactions
in Nature, which are strong interactions, electromagnetic
interactions, weak interactions and gravitational interactions.
All these fundamental interactions can be described by gauge field
theories, which can be regarded as the common nature of all these
fundamental interactions. And we can unify different kinds of
fundamental interactions in the framework of gauge theory. The first
unification of fundamental interactions in human history is the
unification of electric interactions and magnetic interactions,
which is made by Maxwell in 1864. Now, we know that electromagnetic
theory is a $U(1)$ abelian gauge theory. In 1921, H.Weyl tried to
unify electromagnetic interactions and gravitational interactions
in a theory which has local scale invariance\cite{1,2}. Weyl's
original attempt is not successful, however in his great attempt, he
introduced one of the most important concept in modern physics:
gauge transformation and gauge symmetry. After the foundation
of quantum mechanics, V.Fock, H.Weyl and W.Pauli found that quantum
electrodynamics is a $U(1)$ gauge invariant theory\cite{3,4,5}.
\\

In 1954, Yang and Mills proposed non-abelian gauge field theory\cite{6}.
Soon after, non-abelian gauge field theory is applied to elementary
particle theory. In about 1967 and 1968, using the spontaneously
symmetry breaking and Higgs mechanism\cite{7,8,9,10,11,12,13,14},
S.Weinberg and A.Salam proposed the unified electroweak theory,
which is a $SU(2) \times U(1)$ gauge theory\cite{15,16,17}. The
unified electroweak theory is now widely called the standard model.
The predictions of unified electroweak theory have been confirmed
in a large number of experiments, and the intermediate gauge
bosons $W^{\pm}$ and $Z^0$ which are predicted by unified
electroweak model are also found in experiments. 
However, in the traditional standard model, the gravitational
interactions are not considered.  From nineteen
seventies, physicist begin studying Grand Unified theories which try
to unify strong, electromagnetic and weak interactions in a simple
group. At that time, $SU(5)$ model\cite{18,19},
$SO(10)$ model\cite{20,21,22}, $E_6$ model\cite{23,24,25}
and other models\cite{26,27,28} are proposed. In all these attempts,
gravitational interactions are not considred.\\

Gauge treatment of gravity was suggested immediately after the
gauge theory birth itself\cite{29,30,31}. In the traditional 
gauge treatment of gravity, Lorentz group is localized, and the
gravitational field is not represented by gauge
potential\cite{32,33,34}. It is represented by metric field. 
The theory has beautiful mathematical forms, but up to now, 
its renormalizability is not proved. In other words,
it is conventionally considered to be non-renormalizable.
Recently, Wu proposed a new quantum
gauge theory of gravity which is a renormalizable quantum
gravity\cite{35,36}. Based on gauge principle, space-time 
translation group is selected to be the symmetry of 
gravitational interactions, which  appears in a new form 
and is called gravitational gauge group. After
localization of gravitational gauge group, the gravitational 
field appears as the corresponding gauge potential. 
After that work, the unified theory of electromagnetic
interactions and gravitational interactions is discussed\cite{37}.
Then this unified theory is generalized to $SU(N)$ non-abelian
case and the the unification of $SU(N)$ non-abelian
gauge interactions and gravitational interactions, 
which will be called $GSU(N)$ unification theory, 
is discussed\cite{38}. $GSU(N)$ unification theory
is also a renormalizable quantum model\cite{39}. 
Now, using the method propsed in
leteratures \cite{37} and \cite{38}, we will try to
unify four different kinds of fundamental interactions
in Nature in one theory. This unification is based on 
the symmetry of direct or semi-direct product group, which
is the direct extension of the standard model to gravitational
interactions. This unification theory is also perturbatively
renormalzable. 
\\

\section{Gravitational Gauge Field}

First, for the sake of integrity, 
we give a simple introduction to gravitational 
gauge theory and introduce some notations which is used 
in this paper. Details on quantum gauge theory of gravity
can be found in leteratures \cite{35} and \cite{36}. \\

In gauge theory of gravity, the most fundamental 
quantity is gravitaional gauge field $C_{\mu}(x)$.
which is the gauge potential corresponding to gravitational
gauge symmetry. Gauge field $C_{\mu}(x)$ is a vector in 
the corresponding Lie algebra, which, for the sake
of convenience, will be called
gravitational Lie algrbra in this paper. 
So it can expanded as
\be \label{2.10}
C_{\mu}(x) = C_{\mu}^{\alpha} (x) \hat{P}_{\alpha},
\ee
where $C_{\mu}^{\alpha}(x)$ is the component field and
$\hat{P}_{\alpha}$ is the  generator of gravitational
gauge group. 
The gravitational gauge covariant derivative is given by
\be \label{2.9}
D_{\mu} = \partial_{\mu} - i g C_{\mu} (x),
\ee
where $g$ is the gravitational coupling constant.
\\

Matrix $G$ is an important quantity in gauge theory
of gravity, whose definition is
\be \label{2.11}
G = (G_{\mu}^{\alpha}) = ( \delta_{\mu}^{\alpha} - g C_{\mu}^{\alpha} ).
\ee
Its inverse matrix is denoted as $G^{-1}$,
\be \label{2.12}
G^{-1} = \frac{1}{I - gC} = (G^{-1 \mu}_{\alpha}).
\ee
They satisfy the following relations,
\be \label{2.13}
G_{\mu}^{\alpha} G^{-1 \nu}_{\alpha} = \delta_{\mu}^{\nu},
\ee
\be \label{2.14}
G_{\beta}^{-1 \mu} G^{ \alpha}_{\mu} = \delta_{\beta}^{\alpha}.
\ee
It can be proved that
\be \label{2.15}
D_{\mu}= G_{\mu}^{\alpha} \partial_{\alpha}.
\ee
\\

The  field strength of gravitational gauge field is defined by
\be \label{2.16}
F_{\mu\nu} \define \frac{1}{-ig} \lbrack D_{\mu}~~,~~D_{\nu} \rbrack.
\ee
Its explicit expression is
\be \label{2.17}
F_{\mu\nu}(x) = \partial_{\mu} C_{\nu} (x)
-\partial_{\nu} C_{\mu} (x)
- i g C_{\mu} (x) C_{\nu}(x)
+ i g C_{\nu} (x) C_{\mu}(x).
\ee
$F_{\mu\nu}$ is also a vector in gravitational Lie algebra,
\be \label{2.18}
F_{\mu\nu} (x) = F_{\mu\nu}^{\alpha}(x) \cdot \hat{P}_{\alpha},
\ee
where
\be \label{2.19}
F_{\mu\nu}^{\alpha} = \partial_{\mu} C_{\nu}^{\alpha}
-\partial_{\nu} C_{\mu}^{\alpha}
-  g C_{\mu}^{\beta} \partial_{\beta} C_{\nu}^{\alpha}
+  g C_{\nu}^{\beta} \partial_{\beta} C_{\mu}^{\alpha}.
\ee
\\

In order to construct a gravitational gauge invariant
lagrangian, $J(C)$ is an important factor. In this paper,
it will select to be
\be \label{2.26}
J(C) = \sqrt{- {\rm det}( g_{\alpha \beta} ) },
\ee
where
\be \label{2.27}
g_{\alpha \beta} \equiv \eta_{\mu \nu}
(G^{-1})_{\alpha}^{\mu} (G^{-1})_{\beta}^{\nu}.
\ee
\\

\section{Lagrangian and Action }
\setcounter{equation}{0}

Now, we know that there are four kinds of fundamental interactions in
Nature, which are strong interactions, electromagnetic interactions,
weak interactions and gravitational interactions. In the traditional
standard model, the first three fundamental interactions are considered.
In this chapter, we will generalize the standard model to
gravitational interactions. In other words, the new standard model
is a theory that can describe any fundamental physical process
which human kind has already known in Nature.\\

We know that the fundamental particles that we know are
fundamental fermions (such as leptons and quarks),
gauge bosons(such as photon, gluons, gravitons and
intermediate gauge bosons $W^{\pm}$ and $Z^0$), and
possible Higgs bosons. Our goal is to construct a theory
which can describe all possible fundamental interactions
among all these elementary particles. We know that the
symmetry of the traditional standard model is
$SU(3)_c \times SU(2)_L \times U(1)_Y$. $SU(3)_c$ is
the gauge symmetry for strong interactions,
$SU(2)_L \times U(1)_Y$ is the symmetry for electroweak
interactions. The gauge symmetry for gravitational interactions
is gravitational gauge group. If we generalize the standard
model to gravitational interactions, the symmetry group
is
\be \label{4.1}
(SU(3)_c \times SU(2)_L \times U(1)_Y) \otimes_s
Gravitational ~Gauge ~Group.
\ee
This is the symmetry of generalized standard model. \\

Before we construct the lagrangian of the system, let's first
define wave functions of various elementary particles.
According to the Standard Model,
leptons form left-hand doublets and right-hand singlets.
Let's denote
\be \label{4.2}
\psi^{(1)}_L =\left (
\begin{array}{c}
\nu_e  \\
e
\end{array}
\right )_L
~~~,~~~
\psi^{(2)}_L =\left (
\begin{array}{c}
\nu_{\mu}  \\
\mu
\end{array}
\right )_L
~~~,~~~
\psi^{(3)}_L =\left (
\begin{array}{c}
\nu_{\tau}  \\
\tau
\end{array}
\right )_L ,
\ee
\be \label{4.3}
\psi^{(1)}_R=e_R
~~~,~~~
\psi^{(2)}_R=\mu_R
~~~,~~~
\psi^{(3)}_R=\tau_R.
\ee
Neutrinos have no right-hand singlets. The weak hypercharge
for left-hand doublets $\psi^{(i)}_L$ is $-1$ and for right-hand
singlet $\psi^{(i)}_R$ is $-2$. All leptons are $SU(3)_c$ singlet,
so they carry no color charge.
The charges of leptons are given by Gell-Mann-Nishijima rule,
\be \label{4.4}
Q = T_3^L + \frac{Y}{2},
\ee
where $Y$ is the hypercharge and $T_3^L$ is the weak isospin.
In order to define the wave
function for quarks, we have to introduce
Kabayashi-Maskawa mixing matrix first, whose general form is,
\be \label{4.5}
K =
\left (
\begin{array}{ccc}
c_1 & s_1 c_3 & s_1 s_3 \\
-s_1 c_2 & c_1 c_2 c_3 - s_2 s_3 e^{i \delta}
& c_1 c_2 s_3 + s_2 c_3 e^{i \delta}   \\
s_1 s_2 & -c_1 s_2 c_3 -c_2 s_3 e^{i \delta}
& -c_1 s_2 s_3 +c_2 c_3 e^{i \delta}
\end{array}
\right )
\ee
where
\be \label{4.6}
c_i = {\rm cos} \theta_i ~~,~~~~ s_i = {\rm sin} \theta_i ~~~(i=1,2,3)
\ee
and $\theta_i$ are generalized Cabibbo angles. The mixing between
three different quarks $d,s$ and $b$ is given by
\be \label{4.7}
\left (
\begin{array}{c}
d_{\theta} \\
s_{\theta}  \\
b_{\theta}
\end{array}
\right )
= K
\left (
\begin{array}{c}
d\\
s\\
b
\end{array}
\right ).
\ee
Quarks also form left-hand doublets and right-hand singlets,
\be \label{4.8}
q_L^{(1)a} =
\left (
\begin{array}{c}
u_L^a \\
d_{\theta L}^a
\end{array}
\right ) ,~~
q_L^{(2)a} =
\left (
\begin{array}{c}
c_L^a \\
s_{\theta L}^a
\end{array}
\right ),~~
q_L^{(3)a} =
\left (
\begin{array}{c}
t_L^a \\
b_{\theta L}^a
\end{array}
\right ) ,
\ee
\be \label{4.9}
\begin{array}{ccc}
q_u^{(1)a}= u_R^a
& q_u^{(2)a}= c_R^a
& q_u^{(3)a}= t_R^a \\
q_{ \theta d}^{(1)a}= d_{\theta R}^a
& q_{ \theta d}^{(2)a}= s_{\theta R}^a
& q_{ \theta d}^{(3)a}= b_{\theta R}^a,
\end{array}
\ee
where index $a$ is color index.
It is known that left-hand doublets have weak isospin
$\frac{1}{2}$ and weak hypercharge  $\frac{1}{3}$,
right-hand singlets  have no weak isospin, $q_u^{(j)a}$s
have weak hypercharge $\frac{4}{3}$ and $q_{\theta d}^{(j)a}$s
have weak hypercharge $ - \frac{2}{3}$. Charges of quarks are
also given by Gell-Mann-Nishijima rule. \\

The symmetry of system is given by eq.(\ref{4.1}). We can see
that there are four different symmetry group. For every kinds
of symmetry group, there are corresponding gauge fields which
transmit corresponding gauge interactions.
For gauge bosons, gravitational gauge field is also denoted
by $C_{\mu}^{\alpha}$. The coupling constant of gravitational
interactions is still denoted by $g$.
The gluon field is denoted by $A_{\mu}$,
\be \label{4.10}
A_{\mu} =
A_{\mu}^i \frac{\lambda^i}{2}.
\ee
where $\lambda^i$ are Gell-Mann matrix.
Color index is denoted by indexes $a,b,c, \cdots$, so
\be \label{4.11}
\lambda^i = (\lambda^i_{ab}).
\ee
The field strength of gluon field is $A_{\mu\nu}$
\be \label{4.12}
A_{\mu\nu} = ( D_{\mu} A_{\nu} )
 - (D_{\nu} A_{\mu})
- i g_c \lbrack A_{\mu} ~~,~~ A_{\nu} \rbrack
= A^i_{\mu\nu} \frac{\lambda^i}{2},
\ee
where $g_c$ is the coupling constant for strong interactions.
$A^i_{\mu\nu}$ is the component field strength of gluon
field, whose explicit expression is
\be \label{4.13}
A^i_{\mu\nu} = ( D_{\mu} A^i_{\nu} )
 - (D_{\nu} A^i_{\mu})
+ g_c f_{ijk} A^j_{\mu}  A^k_{\nu},
\ee
where $f_{ijk}$ is the structure constant of $SU(3)$ group.
$A_{\mu\nu}$ is not $SU(3)$ gauge covariant field strength.
$SU(3)$ gauge covariant field strength is defined by
\be \label{4.14}
{\mathbb A}_{\mu\nu} =
A_{\mu\nu} + g G^{-1 \lambda}_{\sigma} A_{\lambda}
F_{\sigma}^{\mu\nu}
= {\mathbb A}^i_{\mu\nu} \frac{\lambda^i}{2},
\ee
where
\be \label{4.15}
{\mathbb A}^i_{\mu\nu} =
A^i_{\mu\nu} + g G^{-1 \lambda}_{\sigma} A^i_{\lambda}
F_{\sigma}^{\mu\nu}.
\ee
The $U(1)_Y$ gauge field is denoted by $B_{\mu}$ and the coupling
constant for $U(1)_Y$ gauge interactions is $g'_w$.
The $U(1)_Y$ gauge field strength tensor is $B_{\mu\nu}$
\be \label{4.16}
B_{\mu\nu} = (D_{\mu} B_{\nu}) - (D_{\nu} B_{\mu}).
\ee
The gauge covariant field strength tensor is
\be  \label{4.17}
{\mathbb B}_{\mu\nu} = B_{\mu\nu} + g G^{-1 \lambda}_{\alpha}
B_{\lambda} F^{\alpha}_{\mu\nu},
\ee
The $SU(2)_L$ gauge field is denoted by $W_{\mu}$
\be  \label{4.18}
W_{\mu} =
W^n_{\mu} \frac{\sigma_n}{2},
\ee
where $\sigma_n$ is the Pauli matrix. The $SU(2)_L$ field strength
tensor is $W_{\mu\nu}$,
\be  \label{4.19}
W_{\mu\nu} = (D_{\mu} W_{\nu}) - (D_{\nu} W_{\mu})
- i g_w \lbrack W_{\mu} ~~,~~ W_{\nu} \rbrack
=  W^n_{\mu\nu} \frac{\sigma_n}{2},
\ee
where $g_w$ is the weak coupling constant for $SU(2)_L$ gauge
interactions. $W^n_{\mu\nu}$ is component field strength tensor,
\be  \label{4.20}
W_{\mu \nu}^n = D_{\mu} W_{\nu}^n - D_{\nu} W_{\mu}^n + g_w
\epsilon _{lmn} W_{\mu}^l    W_{\nu}^m,
\ee
where $\epsilon_{lmn}$ is the structure constant of $SU(2)$ group.
The $SU(2)_L$ gauge covariant field strength tensor is
${\mathbb W}_{\mu\nu}$,
\be  \label{4.21}
{\mathbb W}_{\mu\nu} = W_{\mu\nu} + g G^{-1 \lambda}_{\alpha}
W_{\lambda} F^{\alpha}_{\mu\nu}
= {\mathbb W}^n_{\mu\nu} \frac{\sigma_n}{2},
\ee
where
\be  \label{4.22}
{\mathbb W}^n_{\mu\nu} = W^n_{\mu\nu} + g G^{-1 \lambda}_{\alpha}
W^n_{\lambda} F^{\alpha}_{\mu\nu}.
\ee
\\

If there exist Higgs particle in Nature, the Higgs field is
represented by a complex scalar $SU(2)$ doublet,
\be  \label{4.23}
\phi =
\left (
\begin{array}{c}
\phi^{\dagger} \\
\phi^0
\end{array}
\right ).
\ee
The hypercharge of Higgs field $\phi$ is $1$. Higgs field carries no
color charge. \\

The Lagrangian ${\cal L}_0$ that describes four kinds of fundamental
interactions is given by
\be  \label{4.24}
\begin{array}{rcl}
{\cal L}_0 &=&
-\sum_{j=1}^{3} \overline{\psi}_L^{(j)} \gamma ^{\mu}
(D_{\mu}+ \frac{i}{2} g'_w  B_{\mu} -ig_w W_{\mu} ) \psi_L^{(j)} \\
&&\\
&&- \sum_{j=1}^{3}\overline{e}_R^{(j)} \gamma ^{\mu}
(D_{\mu}+ ig'_w  B_{\mu} ) e_R^{(j)}  \\
&&\\
&&-\sum_{j=1}^{3} \overline{q}_L^{(j)a} \gamma^{\mu}
\left( (D_{\mu}-ig_w W_{\mu}- \frac{i}{6}g'_w B_{\mu} )\delta_{ab}
-i g_c A^k_{\mu} (\frac{\lambda^k}{2})_{ab} \right) q_L^{(j)b} \\
&&\\
&&-\sum_{j=1}^{3} \overline{q}_u^{(j)a} \gamma^{\mu}
\left( (D_{\mu}-i \frac{2}{3} g'_w B_{\mu} )\delta_{ab}
-i g_c A^k_{\mu} (\frac{\lambda^k}{2})_{ab} \right) q_u^{(j)b}  \\
&&\\
&& -\sum_{j=1}^{3} \overline{q}_{\theta d}^{(j)a} \gamma^{\mu}
\left( (D_{\mu} + i \frac{1}{3} g'_w B_{\mu} )\delta_{ab}
-i g_c A^k_{\mu} (\frac{\lambda^k}{2})_{ab} \right) q_{\theta d}^{(j)b} \\
&& \\
&&-\frac{1}{4}  \eta^{\mu \rho} \eta^{\nu \sigma}
{\mathbb W}^{n}_{ \mu \nu} {\mathbb W}^n_{\rho \sigma}
-\frac{1}{4} \eta^{\mu \rho} \eta^{\nu \sigma}
{\mathbb B}_{\mu \nu} {\mathbb B}_{\rho \sigma} \\
&& \\
&& -\frac{1}{4} \eta^{\mu \rho} \eta^{\nu \sigma}
{\mathbb A}^i_{\mu \nu } {\mathbb A}^i_{ \rho \sigma }
- \frac{1}{4} \eta^{\mu \rho} \eta^{\nu \sigma} g_{\alpha \beta }
F_{\mu \nu}^{\alpha} F_{\rho \sigma}^{\beta} \\
&&\\
&& -\eta^{\mu\nu} \left\lbrack (D_{\mu}- \frac{i}{2}
g'_w  B_{\mu} -ig_w W_{\mu}) \phi \right\rbrack ^{\dagger}
\cdot \left\lbrack (D_{\nu}- \frac{i}{2}
g'_w  B_{\nu} -ig_w W_{\nu}) \phi \right\rbrack \\
&&\\
&&   - \mu^2 \phi^{\dagger} \phi
+ \lambda (\phi^{\dagger} \phi)^2  \\
&&\\
&& - \sum_{j=1}^{3} f^{(j)}
\left(\overline{e}_R^{(j)} \phi^{\dag} \psi_L^{(j)}
+\overline{\psi}_L^{(j)} \phi  e_R^{(j)}\right)  \\
&&\\
&& -\sum_{j=1}^{3} \left( f_u^{(j)} \overline{q}_L^{(j)a}
\overline{\phi}
q_u^{(j)a} + f_u^{(j) \ast} \overline{q}_u^{(j)a}
\overline{\phi}^{\dag} q_L^{(j)a} \right)   \\
&&\\
&&-\sum_{j,k=1}^{3} \left( f_d^{(jk)} \overline{q}_L^{(j)a} \phi
q_{\theta d}^{(k)a}
+ f_d^{(jk) \ast} \overline{q}_{\theta d}^{(k)a}
\phi^{\dag} q_L^{(j)a} \right),
\end{array}
\ee
where
\be  \label{4.25}
\overline{\phi} = i \sigma_{2} \phi^{\ast} =
\left (
\begin{array}{c}
\phi^{0 \dag} \\
- \phi
\end{array}
\right ).
\ee
The full Lagrangian is given by
\be  \label{4.26}
{\cal L} = J(C) {\cal L}_0,
\ee
and the action of the system is
\be  \label{4.27}
S = \int \d4 x {\cal L} .
\ee
\\

\section{Symmetry of the Model }
\setcounter{equation}{0}

Now, let's discuss symmetry of the system. First, let's study
$SU(3)_c$ symmetry. Denote the $SU(3)_c$ transformation matrix
as $U_3$. Under $SU(3)_c$ transformation, transformations of
various fields and operators are:
\be  \label{4.28}
\psi_L^{(j)} \to \psi_L^{\prime (j)} = \psi_L^{(j)},
\ee
\be  \label{4.29}
e_R^{(j)} \to e_R^{\prime (j)} = e_R^{(j)},
\ee
\be  \label{4.30}
q_L^{(j)a} \to q_L^{\prime (j)a} = U_{3 ab} ~q_L^{(j)b},
\ee
\be  \label{4.31}
q_u^{(j)a} \to q_u^{\prime (j)a} = U_{3 ab} ~q_u^{(j)b},
\ee
\be  \label{4.32}
q_{\theta d}^{(j)a} \to q_{\theta d}^{\prime (j)a}
= U_{3 ab} ~q_{\theta d}^{(j)b},
\ee
\be  \label{4.33}
C_{\mu}^{\alpha}  \to  C_{\mu}^{\prime \alpha}
 = C_{\mu}^{\alpha}
\ee
\be  \label{4.34}
D_{\mu}  \to D'_{\mu} = D_{\mu},
\ee
\be  \label{4.35}
A_{\mu}  \to A'_{\mu} = U_3 A_{\mu} U_3^{-1}
- \frac{1}{i g_c} U_3 ( D_{\mu} U_3^{-1}),
\ee
\be  \label{4.36}
B_{\mu}  \to B'_{\mu} = B_{\mu},
\ee
\be  \label{4.37}
W_{\mu}  \to W'_{\mu} = W_{\mu},
\ee
\be  \label{4.38}
\phi  \to \phi' = \phi,
\ee
\be  \label{4.39}
J(C)  \to J'(C') = J(C).
\ee
According to above transformation rules, gauge field strength
tensors transform as
\be  \label{4.40}
{\mathbb W}_{\mu\nu} \to {\mathbb W}'_{\mu\nu}
= {\mathbb W}_{\mu\nu},
\ee
\be  \label{4.41}
{\mathbb B}_{\mu\nu} \to {\mathbb B}'_{\mu\nu}
= {\mathbb B}_{\mu\nu},
\ee
\be  \label{4.42}
F_{\mu\nu}^{\sigma} \to F_{\mu\nu}^{\prime\sigma}
= F_{\mu\nu}^{\sigma},
\ee
\be  \label{4.43}
A_{\mu\nu} \to A'_{\mu\nu} = U_3 A_{\mu\nu} U_3^{-1}
- \frac{ig}{g_c} F_{\mu\nu}^{\sigma} U_3
(\partial_{\sigma} U_3^{-1} ),
\ee
\be  \label{4.44}
{\mathbb A}_{\mu\nu} \to {\mathbb A}'_{\mu\nu}
= U_3 {\mathbb A}_{\mu\nu} U_3^{-1}.
\ee
Using all these transformation rules, we can prove that the lagrangian
density ${\cal L}$ is invariant. Therefore, the system has strict
local $SU(3)_c$ symmetry. \\

Denote the transformations matrix of $SU(2)_L$ gauge transformation
as $U_2$.Under $SU(2)_L$ transformation, transformations of
various fields are:
\be  \label{4.45}
\psi_L^{(j)} \to \psi_L^{\prime (j)} =  U_2 \psi_L^{(j)},
\ee
\be  \label{4.46}
e_R^{(j)} \to e_R^{\prime (j)} = e_R^{(j)},
\ee
\be  \label{4.47}
q_L^{(j)a} \to q_L^{\prime (j)a} = U_{2} ~q_L^{(j)a},
\ee
\be  \label{4.48}
q_u^{(j)a} \to q_u^{\prime (j)a} =   q_u^{(j)a},
\ee
\be  \label{4.49}
q_{\theta d}^{(j)a} \to q_{\theta d}^{\prime (j)a}
=  q_{\theta d}^{(j)a},
\ee
\be  \label{4.50}
C_{\mu}^{\alpha}  \to  C_{\mu}^{\prime \alpha}
 = C_{\mu}^{\alpha}
\ee
\be  \label{4.51}
D_{\mu}  \to D'_{\mu} = D_{\mu},
\ee
\be  \label{4.52}
A_{\mu}  \to A'_{\mu} =   A_{\mu}  ,
\ee
\be  \label{4.53}
B_{\mu}  \to B'_{\mu} = B_{\mu},
\ee
\be  \label{4.54}
W_{\mu}  \to W'_{\mu} = U_2 W_{\mu} U_2^{-1}
- \frac{1}{i g_W} U_2 (D_{\mu} U_2^{-1} ),
\ee
\be  \label{4.55}
\phi  \to \phi' = U_2 \phi,
\ee
\be  \label{4.56}
J(C)  \to J'(C') = J(C).
\ee
According to above transformation rules, gauge field strength
tensors transform as
\be  \label{4.57}
{\mathbb A}_{\mu\nu} \to {\mathbb A}'_{\mu\nu}
= {\mathbb A}_{\mu\nu},
\ee
\be  \label{4.58}
{\mathbb B}_{\mu\nu} \to {\mathbb B}'_{\mu\nu}
= {\mathbb B}_{\mu\nu},
\ee
\be  \label{4.59}
F_{\mu\nu}^{\sigma} \to F_{\mu\nu}^{\prime\sigma}
= F_{\mu\nu}^{\sigma},
\ee
\be  \label{4.60}
W_{\mu\nu} \to W'_{\mu\nu} = U_2 W_{\mu\nu} U_2^{-1}
- \frac{ig}{g_W} F_{\mu\nu}^{\sigma} U_2
(\partial_{\sigma} U_2^{-1} ),
\ee
\be  \label{4.61}
{\mathbb W}_{\mu\nu} \to {\mathbb W}'_{\mu\nu}
= U_2 {\mathbb W}_{\mu\nu} U_2^{-1}.
\ee
We can prove that the action of the system is invariant under
the above $SU(2)_L$ gauge transformations. \\

Under $U(1)_Y$ gauge transformation, transformations of
various fields are:
\be  \label{4.62}
\psi_L^{(j)} \to \psi_L^{\prime (j)}
=  e^{i \alpha(x) /2} ~\psi_L^{(j)},
\ee
\be  \label{4.63}
e_R^{(j)} \to e_R^{\prime (j)} =  e^{i \alpha(x)}~ e_R^{(j)},
\ee
\be  \label{4.64}
q_L^{(j)a} \to q_L^{\prime (j)a}
= e^{-i \alpha(x) /6} ~q_L^{(j)a},
\ee
\be  \label{4.65}
q_u^{(j)a} \to q_u^{\prime (j)a}
= e^{-2i \alpha(x) /3} ~ q_u^{(j)a},
\ee
\be  \label{4.66}
q_{\theta d}^{(j)a} \to q_{\theta d}^{\prime (j)a}
=  e^{i \alpha(x) /3} q_{\theta d}^{(j)a},
\ee
\be  \label{4.67}
C_{\mu}^{\alpha}  \to  C_{\mu}^{\prime \alpha}
 = C_{\mu}^{\alpha}
\ee
\be  \label{4.68}
D_{\mu}  \to D'_{\mu} = D_{\mu},
\ee
\be  \label{4.69}
A_{\mu}  \to A'_{\mu} =   A_{\mu}  ,
\ee
\be  \label{4.70}
B_{\mu}  \to B'_{\mu} = B_{\mu}
- \frac{1}{g'_W} (D_{\mu} \alpha(x)),
\ee
\be  \label{4.71}
W_{\mu}  \to W'_{\mu} =  W_{\mu} ,
\ee
\be  \label{4.72}
\phi  \to \phi' = e^{-i \alpha(x) /2} \phi,
\ee
\be  \label{4.73}
\bar\phi  \to \bar\phi' = e^{ i \alpha(x) /2} \bar\phi,
\ee
\be  \label{4.74}
J(C)  \to J'(C') = J(C).
\ee
According to above transformation rules, gauge field strength
tensors transform as
\be  \label{4.75}
{\mathbb A}_{\mu\nu} \to {\mathbb A}'_{\mu\nu}
= {\mathbb A}_{\mu\nu},
\ee
\be  \label{4.76}
{\mathbb W}_{\mu\nu} \to {\mathbb W}'_{\mu\nu}
= {\mathbb W}_{\mu\nu},
\ee
\be  \label{4.77}
F_{\mu\nu}^{\sigma} \to F_{\mu\nu}^{\prime\sigma}
= F_{\mu\nu}^{\sigma},
\ee
\be  \label{4.78}
B_{\mu\nu} \to B'_{\mu\nu} =  B_{\mu\nu}
+ \frac{ g}{g'_W} F_{\mu\nu}^{\sigma}
(\partial_{\sigma} \alpha(x) ),
\ee
\be  \label{4.79}
{\mathbb B}_{\mu\nu} \to {\mathbb B}'_{\mu\nu}
=   {\mathbb B}_{\mu\nu} .
\ee
We can also prove that the action of the system is invariant under
the above $U(1)_Y$ gauge transformations.\\

Gravitational gauge transformations of various fields are
\be  \label{4.80}
\psi_L^{(j)} \to \psi_L^{\prime (j)}
=  (\ehat \psi_L^{(j)}),
\ee
\be  \label{4.81}
e_R^{(j)} \to e_R^{\prime (j)} =  (\ehat e_R^{(j)}),
\ee
\be  \label{4.82}
q_L^{(j)a} \to q_L^{\prime (j)a}
= (\ehat q_L^{(j)a}),
\ee
\be  \label{4.83}
q_u^{(j)a} \to q_u^{\prime (j)a}
= (\ehat  q_u^{(j)a}),
\ee
\be  \label{4.84}
q_{\theta d}^{(j)a} \to q_{\theta d}^{\prime (j)a}
=  (\ehat q_{\theta d}^{(j)a}),
\ee
\be  \label{4.85}
C_{\mu}   \to  C_{\mu}^{\prime }
 = \ehat C_{\mu} \ehat^{-1}
 - \frac{1}{ig} \ehat (\partial_{\mu} \ehat^{-1}),
\ee
\be  \label{4.86}
D_{\mu}  \to D'_{\mu} = \ehat D_{\mu} \ehat^{-1},
\ee
\be  \label{4.87}
A_{\mu}  \to A'_{\mu} =  (\ehat A_{\mu})  ,
\ee
\be  \label{4.88}
B_{\mu}  \to B'_{\mu} = (\ehat B_{\mu} ),
\ee
\be  \label{4.89}
W_{\mu}  \to W'_{\mu} = (\ehat  W_{\mu} ),
\ee
\be  \label{4.90}
\phi  \to \phi' = (\ehat \phi) ,
\ee
\be  \label{4.91}
\bar\phi  \to \bar\phi' =(\ehat  \bar\phi),
\ee
\be  \label{4.92}
J(C)  \to J'(C') = J \cdot ( \ehat J(C)),
\ee
\be  \label{4.93}
\eta^{\mu}_{1\alpha}  \to \eta^{\prime\mu}_{1\alpha}
= \Lambda_{\alpha}^{~\alpha_1}
(\ehat \eta^{\mu}_{1\alpha_1}),
\ee
\be  \label{4.94}
\eta_{2\alpha\beta}  \to \eta^{\prime }_{2\alpha\beta}
= \Lambda_{\alpha}^{~\alpha_1} \Lambda_{\beta}^{~\beta_1}
(\ehat \eta_{2\alpha_1 \beta_1}).
\ee
According to above transformation rules, gauge field strength
tensors transform as
\be  \label{4.95}
{\mathbb A}_{\mu\nu} \to {\mathbb A}'_{\mu\nu}
= (\ehat {\mathbb A}_{\mu\nu}),
\ee
\be  \label{4.96}
{\mathbb W}_{\mu\nu} \to {\mathbb W}'_{\mu\nu}
= (\ehat {\mathbb W}_{\mu\nu}),
\ee
\be  \label{4.97}
{\mathbb B}_{\mu\nu} \to {\mathbb B}'_{\mu\nu}
= (\ehat  {\mathbb B}_{\mu\nu}) .
\ee
\be  \label{4.98}
F_{\mu\nu}^{\alpha} \to F_{\mu\nu}^{\prime\alpha}
= \Lambda^{\alpha}_{~\beta} (\ehat F_{\mu\nu}^{\beta}),
\ee
According to these transformations, the lagrangian density
${\cal L}_0$ transforms covariantly,
\be  \label{4.99}
{\cal L}_0 \to {\cal L}'_0 = (\ehat {\cal L_0}).
\ee
So,
\be  \label{4.100}
{\cal L}  \to  {\cal L}' = J \cdot (\ehat {\cal L}).
\ee
Using eq.(2.54), we can prove that action $S$ is invariant
under gravitational gauge transformations,
\be  \label{4.101}
S \to S' = S.
\ee
Therefore, the system has gravitational gauge symmetry. \\

Now, as a whole, we discuss
$$
(SU(3)_c \times SU(2)_L \times U(1)_Y) \otimes_s
Gravitational ~Gauge ~Group
$$
gauge symmetry. In order to do this, we need define generator
operators. The generator operators of $SU(3)_c$ group are denoted
by $\hat T_{3j}$. The $SU(3)_c$ transformation operator $\hat U_3$
is defined by
\be  \label{4.102}
\hat U_3 = e^{ -i \alpha^j \hat T_{3j} }.
\ee
Matrix $U_3$ is defined by
\be  \label{4.103}
U_3 = e^{-i \alpha^j \lambda^j /2 }.
\ee
When $\hat T_{3j}$ acts on fields, it will becomes
the corresponding representation matrix of generators. So,
\be  \label{4.104}
\hat T_{3i} \psi_L^{(j)} = 0,
\ee
\be  \label{4.105}
\hat T_{3i} e_R^{(j)} = 0,
\ee
\be  \label{4.106}
\hat T_{3i} q_L^{(j)a} =
\left( \frac{\lambda^i}{2} \right)_{ab}
q_L^{(j)b},
\ee
\be  \label{4.107}
\hat T_{3i} q_u^{(j)a} =
\left( \frac{\lambda^i}{2} \right)_{ab}
q_u^{(j)b},
\ee
\be  \label{4.108}
\hat T_{3i} q_{\theta d}^{(j)a} =
\left( \frac{\lambda^i}{2} \right)_{ab}
q_{\theta d}^{(j)b},
\ee
\be  \label{4.109}
\hat T_{3i} \phi = 0,
\ee
\be  \label{4.110}
\lbrack \hat T_{3i} ~~,~~ C_{\mu}^{\alpha} \rbrack = 0,
\ee
\be  \label{4.111}
\lbrack \hat T_{3i} ~~,~~ A_{\mu} \rbrack
= \left\lbrack \frac{\lambda^i}{2}
~~,~~A_{\mu}  \right\rbrack,
\ee
\be  \label{4.112}
\lbrack \hat T_{3i} ~~,~~ B_{\mu} \rbrack = 0,
\ee
\be  \label{4.113}
\lbrack \hat T_{3i} ~~,~~ W_{\mu} \rbrack = 0,
\ee
\be  \label{4.114}
\lbrack \hat T_{3i} ~~,~~ F_{\mu\nu}^{\sigma} \rbrack = 0,
\ee
\be  \label{4.115}
\lbrack \hat T_{3i} ~~,~~ {\mathbb A}_{\mu\nu} \rbrack
= \left\lbrack \frac{\lambda^i}{2}
~~,~~{\mathbb A}_{\mu\nu}  \right\rbrack,
\ee
\be  \label{4.116}
\lbrack \hat T_{3i} ~~,~~ {\mathbb B}_{\mu\nu} \rbrack =0,
\ee
\be  \label{4.117}
\lbrack \hat T_{3i} ~~,~~ {\mathbb W}_{\mu\nu} \rbrack =0.
\ee
The generator operators of $SU(2)_L$ group are denoted
by $\hat T_{2l}$. The $SU(2)_L$ transformation operator $\hat U_2$
is defined by
\be  \label{4.118}
\hat U_2 = e^{ -i \alpha^l \hat T_{2l} }.
\ee
Matrix $U_2$ is defined by
\be  \label{4.119}
U_2 = e^{-i \alpha^l \sigma_l /2 }.
\ee
When $\hat T_{2l}$ acts on fields, it will becomes
the corresponding representation matrix of generators. So,
\be  \label{4.120}
\hat T_{2l} \psi_L^{(j)} = \frac{\sigma_l}{2} \psi_L^{(j)}
\ee
\be  \label{4.121}
\hat T_{2l} e_R^{(j)} = 0,
\ee
\be  \label{4.122}
\hat T_{2l} q_L^{(j)a} =
 \frac{\sigma_l}{2} q_L^{(j)a},
\ee
\be  \label{4.123}
\hat T_{2l} q_u^{(j)a} = 0,
\ee
\be  \label{4.124}
\hat T_{2l} q_{\theta d}^{(j)a} = 0,
\ee
\be  \label{4.125}
\hat T_{2l} \phi = \frac{\sigma_l}{2} \phi
\ee
\be  \label{4.126}
\lbrack \hat T_{2l} ~~,~~ C_{\mu}^{\alpha} \rbrack = 0,
\ee
\be  \label{4.127}
\lbrack \hat T_{2l} ~~,~~ A_{\mu} \rbrack =0,
\ee
\be  \label{4.128}
\lbrack \hat T_{2l} ~~,~~ B_{\mu} \rbrack = 0,
\ee
\be  \label{4.129}
\lbrack \hat T_{2l} ~~,~~ W_{\mu} \rbrack =
\left\lbrack \frac{\sigma_l}{2} ~~,~~W_{\mu} \right\rbrack,
\ee
\be  \label{4.130}
\lbrack \hat T_{2l} ~~,~~ F_{\mu\nu}^{\sigma} \rbrack = 0,
\ee
\be  \label{4.131}
\lbrack \hat T_{2l} ~~,~~ {\mathbb A}_{\mu\nu} \rbrack =0,
\ee
\be  \label{4.132}
\lbrack \hat T_{2l} ~~,~~ {\mathbb B}_{\mu\nu} \rbrack =0,
\ee
\be  \label{4.133}
\lbrack \hat T_{2l} ~~,~~ {\mathbb W}_{\mu\nu} \rbrack =
\left\lbrack \frac{\sigma_l}{2} ~~,~~
{\mathbb W}_{\mu\nu} \right\rbrack .
\ee
The generator operators of $U(1)_Y$ group are denoted
by $\hat T_1$. $2\hat T_1$ is the hypercharge operator.
The $U(1)_Y$ transformation operator $\hat U_1$
is defined by
\be  \label{4.134}
\hat U_1 = e^{ -i \alpha  \hat T_1 }.
\ee
Matrix $U_1$ is defined by
\be  \label{4.135}
U_1 = e^{-i \alpha }.
\ee
When $2 \hat T_1$ acts on fields, it will becomes the hypercharge
of the corresponding fields. So,
\be  \label{4.136}
\hat T_{1} \psi_L^{(j)} = - \frac{1}{2} \psi_L^{(j)}
\ee
\be  \label{4.137}
\hat T_{1} e_R^{(j)} = - e_R^{(j)}
\ee
\be  \label{4.138}
\hat T_{1} q_L^{(j)a} =
 \frac{1}{6} q_L^{(j)a},
\ee
\be  \label{4.139}
\hat T_{1} q_u^{(j)a} = \frac{2}{3}  q_u^{(j)a},
\ee
\be  \label{4.140}
\hat T_{1} q_{\theta d}^{(j)a} = -\frac{1}{3} q_{\theta d}^{(j)a}
\ee
\be  \label{4.141}
\hat T_{1} \phi = \frac{1}{2} \phi
\ee
\be  \label{4.142}
\lbrack \hat T_{1} ~~,~~ C_{\mu}^{\alpha} \rbrack = 0,
\ee
\be  \label{4.143}
\lbrack \hat T_{1} ~~,~~ A_{\mu} \rbrack =0,
\ee
\be  \label{4.144}
\lbrack \hat T_{1} ~~,~~ B_{\mu} \rbrack = B_{\mu},
\ee
\be  \label{4.145}
\lbrack \hat T_{1} ~~,~~ W_{\mu} \rbrack = 0,
\ee
\be  \label{4.146}
\lbrack \hat T_{1} ~~,~~ F_{\mu\nu}^{\sigma} \rbrack = 0,
\ee
\be  \label{4.147}
\lbrack \hat T_{1} ~~,~~ {\mathbb A}_{\mu\nu} \rbrack =0,
\ee
\be  \label{4.148}
\lbrack \hat T_{1} ~~,~~ {\mathbb B}_{\mu\nu} \rbrack
= {\mathbb B}_{\mu\nu} ,
\ee
\be  \label{4.149}
\lbrack \hat T_{1} ~~,~~ {\mathbb W}_{\mu\nu} \rbrack = 0.
\ee
Different generator operators act on different spaces, so they
commute each other,
\be  \label{4.150}
\lbrack \hat T_{1} ~~,~~\hat T_{2l} \rbrack = 0,
\ee
\be  \label{4.151}
\lbrack \hat T_{1} ~~,~~ \hat T_{3i}\rbrack= 0,
\ee
\be  \label{4.152}
\lbrack \hat T_{1} ~~,~~ \hat P_{\alpha} \rbrack= 0,
\ee
\be  \label{4.153}
\lbrack \hat T_{2l} ~~,~~ \hat T_{3i} \rbrack= 0,
\ee
\be  \label{4.154}
\lbrack \hat T_{2l} ~~,~~ \hat P_{\alpha} \rbrack= 0,
\ee
\be  \label{4.155}
\lbrack \hat T_{3i} ~~,~~ \hat P_{\alpha} \rbrack= 0.
\ee
As we have mention before, what generators commute each other
does not means that group elements commute each other. \\

A general element of semi-direct product group
$$
(SU(3)_c \times SU(2)_L \times U(1)_Y) \otimes_s
Gravitational ~Gauge ~Group
$$
is denoted by $g(x)$. It can be proved that the $g(x)$ can be
written into the following form
\be  \label{4.156}
g(x) = \ehat \hat U_1 \hat U_2 \hat U_3.
\ee
Define quark color triplet states,
\be \label{4.157}
q_{L}^{(j)} =
\left (
\begin{array}{c}
 q_{L}^{(j)1} \\
 q_{L}^{(j)2}  \\
 q_{L}^{(j)3}
\end{array}
\right ),
\ee
\be \label{4.158}
q_{u}^{(j)} =
\left (
\begin{array}{c}
 q_{u}^{(j)1} \\
 q_{u}^{(j)2}  \\
 q_{u}^{(j)3}
\end{array}
\right ),
\ee
\be \label{4.159}
q_{\theta d}^{(j)} =
\left (
\begin{array}{c}
 q_{\theta d}^{(j)1} \\
 q_{\theta d}^{(j)2}  \\
 q_{\theta d}^{(j)3}
\end{array}
\right ).
\ee
Then, we have the following relations
\be  \label{4.160}
\hat T_{3j} q_{L}^{(j)}
= \frac{\lambda^j}{2} q_L^{(j)} ,
\ee
\be  \label{4.161}
\hat T_{3j} q_{u}^{(j)}
= \frac{\lambda^j}{2} q_u^{(j)} ,
\ee
\be  \label{4.162}
\hat T_{3j} q_{\theta d}^{(j)}
= \frac{\lambda^j}{2} q_{\theta d}^{(j)} .
\ee
Under gauge transformations of semi-direct product group,
various fields and operators transform as
\be  \label{4.163}
\psi_L^{(j)} \to \psi_L^{\prime (j)}
=  (g(x) \psi_L^{(j)}),
\ee
\be  \label{4.164}
e_R^{(j)} \to e_R^{\prime (j)} =  (g(x) e_R^{(j)}),
\ee
\be  \label{4.165}
q_L^{(j)a} \to q_L^{\prime (j) }
= (g(x) q_L^{(j) }),
\ee
\be  \label{4.166}
q_u^{(j)a} \to q_u^{\prime (j) }
= (g(x)  q_u^{(j) }),
\ee
\be  \label{4.167}
q_{\theta d}^{(j)a} \to q_{\theta d}^{\prime (j) }
=  (g(x) q_{\theta d}^{(j) }),
\ee
\be  \label{4.168}
C_{\mu}   \to  C_{\mu}^{\prime }
 = \ehat C_{\mu} \ehat^{-1}
 - \frac{1}{ig} \ehat (\partial_{\mu} \ehat^{-1}),
\ee
\be  \label{4.169}
D_{\mu}  \to D'_{\mu} = \ehat D_{\mu} \ehat^{-1},
\ee
\be  \label{4.170}
F_{\mu\nu}  \to F'_{\mu\nu}
=  \ehat F_{\mu\nu} \ehat^{-1}  ,
\ee
\be  \label{4.171}
A_{\mu}  \to A'_{\mu}
= g(x) \left\lbrack   A_{\mu} - \frac{1}{i g_c}
(D_{\mu} U_3^{-1}) U_3 \right\rbrack g^{-1} (x) ,
\ee
\be  \label{4.172}
W_{\mu}  \to W'_{\mu}
= g(x) \left\lbrack   W_{\mu} - \frac{1}{i g_W}
(D_{\mu} U_2^{-1}) U_2 \right\rbrack g^{-1} (x) ,
\ee
\be  \label{4.173}
B_{\mu}  \to B'_{\mu}
= g(x) \left\lbrack   B_{\mu} - \frac{1}{i g'_W}
(D_{\mu} U_1^{-1}) U_1 \right\rbrack g^{-1} (x) ,
\ee
\be  \label{4.174}
\phi  \to \phi' = (g(x) \phi) ,
\ee
\be  \label{4.175}
J(C)  \to J'(C') = J \cdot ( \ehat J(C)),
\ee
\be  \label{4.176}
\eta^{\mu}_{1\alpha}  \to \eta^{\prime\mu}_{1\alpha}
= \Lambda_{\alpha}^{~\alpha_1}
g(x) \eta^{\mu}_{1\alpha_1} g^{-1}(x),
\ee
\be  \label{4.177}
\eta_{2\alpha\beta}  \to \eta^{\prime }_{2\alpha\beta}
= \Lambda_{\alpha}^{~\alpha_1} \Lambda_{\beta}^{~\beta_1}
g(x) \eta_{2\alpha_1 \beta_1} g^{-1}(x),
\ee
\be  \label{4.178}
{\mathbb A}_{\mu\nu} \to {\mathbb A}'_{\mu\nu}
= g(x) {\mathbb A}_{\mu\nu} g^{-1}(x),
\ee
\be  \label{4.179}
{\mathbb W}_{\mu\nu} \to {\mathbb W}'_{\mu\nu}
= g(x) {\mathbb W}_{\mu\nu} g^{-1}(x),
\ee
\be  \label{4.180}
{\mathbb B}_{\mu\nu} \to {\mathbb B}'_{\mu\nu}
= g(x)  {\mathbb B}_{\mu\nu} g^{-1}(x) .
\ee
It can be proved that the action of the system has strict
local gauge symmetry of semi-direct product group.\\

\section{Spontaneously Symmetry Breaking }
\setcounter{equation}{0}

It is known that, $SU(3)_c$ color symmetry and gravitational gauge
symmetry are strict symmetry. $SU(2)_L \times U(1)_Y$ symmetry
are not strict symmetry, which is broken to $U(1)_Q$ symmetry.
Now, let's discuss spontaneously symmetry breaking of the system.
The potential of Higgs field is
\be  \label{4.181}
- \mu^2 \phi^{\dag} \phi
+ \lambda (\phi^{\dag} \phi  )^2.
\ee
If,
\be  \label{4.182}
\mu^2 > 0,~~~~~~
\lambda>0,
\ee
the symmetry of vacuum will be spontaneously broken. Suppose that
the vacuum expectation value of neutral Higgs field is non-zero,
that is
\be  \label{4.183}
\langle \phi \rangle_0 =
\left(
\ba{c}
0\\
\frac{v}{\sqrt{2}}
\ea
\right),
\ee
where,
\be  \label{4.184}
v = \sqrt{\frac{\mu^2}{\lambda}  }.
\ee
After a local $SU(2)_L$ gauge transformation, we can select
the Higgs field $\phi(x)$ as,
\be  \label{4.185}
\langle \phi \rangle_0 = \frac{1}{\sqrt{2}}
\left(
\ba{c}
0\\
v + \varphi(x)
\ea
\right).
\ee
After symmetry breaking, the Higgs potential becomes,
\be  \label{4.186}
V( \varphi ) =
\mu^2 \varphi^2 + \lambda v \varphi^3
+\frac{\lambda}{4} \varphi^4
- \frac{\mu^4}{4 \lambda},
\ee
from which we know that the mass of Higgs field is $2 \mu^2$.\\

Define
\be  \label{4.187}
W_{\mu}^{\pm} \define
\frac{1}{\sqrt{2}} \left(
W_{\mu}^1 \mp i W_{\mu}^2,
\right)
\ee
\be  \label{4.188}
A^e_{\mu} \define \cos \theta_W B_{\mu}
+ \sin \theta_W W_{\mu}^3,
\ee
\be  \label{4.189}
Z_{\mu} \define \sin \theta_W B_{\mu}
- \cos \theta_W W_{\mu}^3,
\ee
where
\be  \label{4.190}
{\rm tg} \theta_W = \frac{g'_W}{g_W}.
\ee
$A^e_{\mu}$ in eq.(\ref{4.188}) is the electromagnetic field.
Define two mass parameters,
\be  \label{4.191}
m_W = \frac{1}{2} g_W v,
\ee
\be  \label{4.192}
m_Z = \frac{1}{2} \sqrt{g_W^2 + g_W^{\prime 2}} v.
\ee
It is known that, in the standard model, $m_W$ is the mass of
$W^{\pm}$ bosons and $m_Z$ is the mass of $Z$ bosons.
The coupling constant of electromagnetic interactions is
denoted by e,
\be  \label{4.193}
{\rm e} \define g'_W \cos \theta_W
= \frac{g_W g'_w}{\sqrt{g_W^2 + g_W^{\prime 2} }}.
\ee
The current of strong interactions are denoted by
$J^{\mu}_{c i}$,
\be  \label{4.194}
J^{\mu}_{c i} = i \left(
\bar u \gamma^{\mu} \frac{\lambda^i}{2} u
+ \bar c \gamma^{\mu} \frac{\lambda^i}{2} c
+ \bar t \gamma^{\mu} \frac{\lambda^i}{2} t
+ \bar d \gamma^{\mu} \frac{\lambda^i}{2} d
+ \bar s \gamma^{\mu} \frac{\lambda^i}{2} s
+ \bar b \gamma^{\mu} \frac{\lambda^i}{2} b
\right).
\ee
The current of electromagnetic interactions is
\be  \label{4.195}
\ba{rcl}
J^{\mu}_{em} & = & i \left(
- \bar e \gamma^{\mu} e
- \bar \mu \gamma^{\mu} \mu
- \bar \tau \gamma^{\mu} \tau
+ \frac{2}{3} \bar u \gamma^{\mu} u
+ \frac{2}{3} \bar c \gamma^{\mu} c \right.  \\
&&\\
&&\left.
   + \frac{2}{3} \bar t \gamma^{\mu} t
   - \frac{1}{3} \bar d \gamma^{\mu} d
   - \frac{1}{3} \bar s \gamma^{\mu} s
   - \frac{1}{3} \bar b \gamma^{\mu} b
\right).
\ea
\ee
The currents for weak interactions are
\be  \label{4.196}
\ba{rcl}
J^{\mu -}_{W} & = & \frac{i}{2 \sqrt{2}} \left\lbrack
 \bar \nu_e \gamma^{\mu} (1 + \gamma_5) e
+ \bar \nu_{\mu} \gamma^{\mu} (1 + \gamma_5) \mu
+ \bar \nu_{\tau} \gamma^{\mu} (1 + \gamma_5) \tau  \right.\\
&&  \\
&&\left.
   + \bar u \gamma^{\mu} (1 + \gamma_5) d_{\theta}
   + \bar c \gamma^{\mu} (1 + \gamma_5) s_{\theta}
   + \bar t \gamma^{\mu} (1 + \gamma_5) b_{\theta}
\right),
\ea
\ee
\be  \label{4.197}
\ba{rcl}
J^{\mu -}_{W} & = & \frac{i}{2 \sqrt{2}} \left\lbrack
  \bar e \gamma^{\mu} (1 + \gamma_5) \nu_e
+ \bar \mu \gamma^{\mu} (1 + \gamma_5) \nu_{\mu}
+ \bar \tau \gamma^{\mu} (1 + \gamma_5) \nu_{\tau}  \right.\\
&&  \\
&&\left.
   + \bar d_{\theta} \gamma^{\mu} (1 + \gamma_5) u
   + \bar s_{\theta} \gamma^{\mu} (1 + \gamma_5) c
   + \bar b_{\theta} \gamma^{\mu} (1 + \gamma_5) t
\right),
\ea
\ee
\be  \label{4.198}
J_z^{\mu} = J_3^{\mu} - \sin ^2 \theta_W J^{\mu}_{em},
\ee
where
\be  \label{4.199}
\ba{rcl}
J^{\mu  }_{3} & = & \frac{i}{2 } \left(
+ \bar \nu_{eL} \gamma^{\mu}  \nu_{eL}
+ \bar \nu_{\mu L} \gamma^{\mu}  \nu_{\mu L}
+ \bar \nu_{\tau L} \gamma^{\mu}  \nu_{\tau L}
- \bar e_L \gamma^{\mu}  e_L
- \bar \mu_L \gamma^{\mu}  \mu_L
- \bar \tau_L \gamma^{\mu}  \tau_L  \right.  \\
&&\\
&& \left.
+ \bar u_L \gamma^{\mu}  u_L
+ \bar c_L \gamma^{\mu}  c_L
+ \bar t_L \gamma^{\mu}  t_L
- \bar d_L \gamma^{\mu}  d_L
- \bar s_L \gamma^{\mu}  s_L
- \bar b_L \gamma^{\mu}  b_L
\right).
\ea
\ee
The current for gravitational interactions is
\be  \label{4.200}
\ba{rcl}
J^{\mu  }_{g \alpha} & = &
 \bar \nu_{eL} \gamma^{\mu} \partial_{\alpha} \nu_{eL}
+ \bar \nu_{\mu L} \gamma^{\mu} \partial_{\alpha}  \nu_{\mu L}
+ \bar \nu_{\tau L} \gamma^{\mu} \partial_{\alpha}  \nu_{\tau L}
+ \bar e_L \gamma^{\mu} \partial_{\alpha}  e_L  \\
&&\\
&&
+ \bar \mu_L \gamma^{\mu} \partial_{\alpha}  \mu_L
+ \bar \tau_L \gamma^{\mu} \partial_{\alpha}  \tau_L
+ \bar u_L \gamma^{\mu} \partial_{\alpha}  u_L
+ \bar c_L \gamma^{\mu} \partial_{\alpha}  c_L  \\
&&\\
&&
+ \bar t_L \gamma^{\mu} \partial_{\alpha}  t_L
+ \bar d_L \gamma^{\mu} \partial_{\alpha}  d_L
+ \bar s_L \gamma^{\mu} \partial_{\alpha}  s_L
+ \bar b_L \gamma^{\mu} \partial_{\alpha}  b_L .
\ea
\ee
Denote
\be  \label{4.201}
J_{\varphi} = \frac{1}{v} (
m_e \bar e e
+ m_{\mu} \bar \mu \mu
+ m_{\tau} \bar \tau \tau
+ m_u \bar u u
+ m_c \bar c c
+ m_t \bar t t
+ m_d \bar d d
+ m_s \bar s s
+ m_b \bar b b
).
\ee
Then lagrangian density ${\cal L}_0$ can be written into the
following form,
\be  \label{4.202}
\ba{rcl}
{\cal L}_0 & = &
- \bar \nu_{eL} \gamma^{\mu} \partial_{\mu} \nu_{eL}
- \bar \nu_{\mu L} \gamma^{\mu} \partial_{\mu}  \nu_{\mu L}
- \bar \nu_{\tau L} \gamma^{\mu} \partial_{\mu}  \nu_{\tau L}
- \bar e  ( \gamma^{\mu} \partial_{\mu} + m_e )  e   \\
&&\\
&&
- \bar \mu  ( \gamma^{\mu} \partial_{\mu} + m_{\mu} ) \mu
- \bar \tau ( \gamma^{\mu} \partial_{\mu} + m_{\tau} )  \tau
- \bar u  (\gamma^{\mu} \partial_{\mu} + m_u ) u
- \bar c  (\gamma^{\mu} \partial_{\mu} + m_c ) c   \\
&&\\
&&
- \bar t  (\gamma^{\mu} \partial_{\mu} + m_t ) t
- \bar d  (\gamma^{\mu} \partial_{\mu} + m_d ) d
- \bar s  (\gamma^{\mu} \partial_{\mu} + m_s ) s
- \bar b  (\gamma^{\mu} \partial_{\mu} + m_b ) b   \\
&&\\
&&
+ g_c J_{c i}^{\mu} A_{\mu}^i
+ g_W J_W^{\mu -} W_{\mu}^+
+ g_W J_W^{\mu +} W_{\mu}^-
- \frac{g_W}{\cos \theta_W} J_z^{\mu} Z_{\mu}
+ {\rm e} J^{\mu}_{em} A^e_{\mu}
+ g J^{\mu}_{g \alpha} C_{\mu}^{\alpha}  \\
&&\\
&&
- \frac{1}{2} \eta^{\mu\nu} (D_{\mu} \varphi)(D_{\nu}\varphi)
- \frac{1}{4} g_W^2 \eta^{\mu\nu}
W_{\mu}^+  W_{\nu}^- (2 v \varphi + \varphi^2)\\
&&\\
&&
-\frac{1}{8} (g_W^2 + g_W^{\prime 2}) \eta^{\mu\nu}
Z_{\mu} Z_{\nu} (2 v \varphi + \varphi^2)
- v(\varphi) - J_{\varphi} \phi  \\
&&\\
&& - \frac{1}{4} \eta^{\mu\rho} \eta^{\nu\sigma}
{\mathbb W}^n_{\mu\nu} {\mathbb W}^n_{\rho\sigma}
- m_W^2 \eta^{\mu\nu} W^+_{\mu} W^-_{\nu}
- \frac{1}{2} m^2_Z \eta^{\mu\nu} Z_{\mu} Z_{\nu} \\
&&\\
&&
- \frac{1}{4} \eta^{\mu\rho} \eta^{\nu\sigma}
{\mathbb B}_{\mu\nu} {\mathbb B}_{\rho\sigma}
- \frac{1}{4} \eta^{\mu\rho} \eta^{\nu\sigma}
{\mathbb A}^i_{\mu\nu} {\mathbb A}^i_{\rho\sigma}
- \frac{1}{4} \eta^{\mu\rho} \eta^{\nu\sigma} \eta_{2\alpha\beta}
F^{\alpha}_{\mu\nu} F^{\beta}_{\rho\sigma}.
\ea
\ee
\\

According to above discussions, before spontaneously symmetry
breaking,  the system has
$$
(SU(3)_c \times SU(2)_L \times U(1)_Y) \otimes_s
Gravitational ~Gauge ~Group
$$
gauge symmetry. After spontaneously symmetry breaking, the system
has
$$
(SU(3)_c  \times U(1)_Q) \otimes_s
Gravitational ~Gauge ~Group
$$
gauge symmetry.
Four different kinds of fundamental interactions
are unified in the same lagrangian. So, the lagrangian density
${\cal L}$ which is given by eq.(\ref{4.24}) and eq.(\ref{4.26})
can be used to calculate any fundamental interaction process
in Nature. \\

\section{Summary and Discussions}

In this paper, we have studied unifications of four
kinds of fundamental interactions in Nature, which is
unified in the direct or semi-direct product group.
This model is the direct extension of the standard
model to gravitational interactions. \\

It is know that the real spirit of the traditional
unification is to try
to unify different kinds of fundamental interactions
in a single simple Lie-group, where only one coupling
constant is used in the unification. But now, we find
that it is impossible to unify gravitational interactions
with other fundamental interactions in a theory which
has only one independent parameter for 
coupling constant. The reason is that,
the generators of gravitational gauge group have
mass dimension, while the generators of ordinary
$SU(N)$ group are dimensionless. If we use only one
coupling constant in a unified theory, we will need
another independent parameters to balance the difference
of the dimensions of generators. It means that we 
really use two independent parameters for coupling constant,
which is equivalent to the theory with two independent 
coupling constants. Therefore, any unification theory
of gravity must need at least two independent parameters
for coupling constant. In other words,
it is impossible to unify four
kinds of fundamental interactions in a simple group in which
only one independent parameter for coupling constant is used. \\

Quantization of the unified theory of fundamental
interactions can be performed in the path integral
method. In order to keep the unitarity of the
S-matrix, ghost fields for gravitational gauge
symmetry, $SU(2)_L$ symmetry and $SU(3)_c$ color
symmetry are needed\cite{40}. We can obtain
Feynmann rules for various interaction vertices and
propagaters for matter fields, gauge fields and
ghost fields. After obtain Feynmann rules, we can
calculate Feynmann dagrams for various interaction
processes and determine finite quantum modificaitons
in the classical theory of gravity.
\\

It is known that quantum gauge theory of gravity
is a perturbatively renormalizable quantum theory.
Detailed proof on its renormalizability can be found 
in literature \cite{35,36}. The foundations of the
unified theory of fundamental interactions which is 
discussed in this paper is the $GSU(N)$ unification theory 
which is  discussed in literature \cite{37,38}. Because the
theory has strict $GSU(N)$ symmetry, it is believed
that the theory is also renormalizable. Detailed study
shows that the strict local $GSU(N)$ symmetry gives out
two sets of generalized BRST transformations which will
give out two sets of generalized Ward-Takahashi identities.
Using these two sets of generalized Ward-Takahashi 
identities, we can determine the general form of the 
divergent part of the generating functional of regular 
vertex. After the divergent part of the generating 
functional is cancelled by counterterm, the renormalized 
generating functional is just a scale transformation of the
non-renormalized generating functional. Comparing the 
renormalized theory with the non-renormalized theory, 
we will find that the only effect of the renormalization 
is to redefine the fields, coupling constants and some 
other parameters of the original theory. Only a few new 
parameters are introduced in the renormalization, so 
the $GSU(N)$ unification theory is indeed a 
renormalizable quantum theory\cite{39}. 
Detailed proof on its renormalizability is quite
complicated mathematically, which will be discussed
in a separate paper. Unified theory of fundamental 
interactions is just an application of the $GSU(N)$ 
unification theory to fundamental interactions in Nature.
Because the $GSU(N)$ unification theory is renormalizable
and from the standard model, we know that the spontaneously 
symmetry breaking does not affect the renormalizability
of the theory, the unified theory of fundamental interactions 
which is discussed in this paper is a renormalizable quantum
theory. Detailed and strict proof on its renormalizability
will be found in our future work\cite{40}. 
This is the first renormalizable quantum model 
which contains four different kinds of fundamental 
interactions in Nature. In the future, we will use
this model to study quantum effects of gravitational 
interactions and new effects which are originated from 
the non-Abelian nature of $GSU(N)$ group and lead
to direct coupling between gravitational gauge field and
ordinary gauge fields(such as electromagnetic field, 
intermediate gauge field and gluon field).  
\\


\begin{thebibliography}{99}

\bibitem{1} H.Weyl, in {\it Raum. Zeit. Materia, } 3rd ed.
    (Springer-Verlag, Berlin, 1920)
\bibitem{2} H.Weyl, {\it Gravitation and Elektrizitat.}
    Sitzungsber. Akademie der Wissenschaften Berlin,
    465-480(1918). Siehe auch die {\it Gesammelten Abhandlungen.}
    6 Vols. Ed. K.Chadrasekharan, Springer-Verlag.
\bibitem{3} V.Fock, Zeit. f. Physik, {\bf 39} (1927) 226.
\bibitem{4} H.Weyl, Zeit. f. Physik, {\bf 56} (1929) 330.
\bibitem{5} W.Pauli, {\it Handbuch der physik, Geiger and
    scheel, 2, Aufl.,} Vol. 24, Teil {\bf 1} pp. 83-272.
\bibitem{6} C.N.Yang, R.L.Mills, Phys Rev {\bf 96} (1954) 191 .
\bibitem{7} J.Goldstone, Nov.Cim. {\bf 19} (1961) 154.
\bibitem{8} Y.Nambu and G.Jona-Lasinio, Phys.Rev. {\bf 122}
        (1961) 34.
\bibitem{9} J.Goldstone, A.Salam and S.Weinberg, Phys.Rev.
        {\bf 127} (1962) 965.
\bibitem{10} P.W.Higgs, Phys.Lett. {\bf 12} (1964) 132.
\bibitem{11} F.Englert and R.Brout, Phys.Rev.Lett.
        {\bf 13} (1964) 321.
\bibitem{12} G.S.Guralnik, C.R.Hagen and T.W.B.Kibble,
        Phys.Rev.Lett. (1964) 585.
\bibitem{13} P.W.Higgs, Phys.Rev. {\bf 145} (1966) 1156.
\bibitem{14} S.Weinberg, Phys.Rev. {\bf D7} (1973) 1068.
\bibitem{15} S.Glashow, Nucl.Phys. {\bf 22}(1961) 579 .
\bibitem{16} S.Weinberg, Phys.Rev.Lett. {\bf 19} (1967) 1264 .
\bibitem{17} A.Salam, in Elementary Particle Theory, eds.N.Svartho .
    Forlag, Stockholm,1968).
\bibitem{18} H.Georgi and S.L.Glashow, Phys.Rev.Lett.
        {\bf 32} (1974) 438.
\bibitem{19} A.J.Buras, J.Ellis, M.K.Gaillard and D.V.Nanopoulos,
        Nucl.Phys. {\bf B135} (1978) 66.
\bibitem{20} H.Georgi, {\it Particles and Fields} 1974, ed. C.E.Carlson
        (AIP, NY, 1975) p.573.
\bibitem{21} H.Fritzsch and P.Minkowski, Ann.Phys. {\bf 93} (1975) 193.
\bibitem{22} M.Machacek, Nucl.Phys. {\bf B159} (1979) 37.
\bibitem{23} F.Gursey and M.Serdaroglu, Lett.Nuo.Cim. {\bf 21} (1978) 28.
\bibitem{24} Y.Achiman and B.Stech, Phys.Lett. {\bf 77B} (1978) 389.
\bibitem{25} R.Barbieri and D.V.Nanopoulos, Phys.Lett.
        {\bf 91B} (1980) 369.
\bibitem{26} J.C.Pati and A.Salam, Phys.Rev.{\bf D8} (1973) 1240;
        Phys.Rev.Lett. {\bf 36} (1976) 1229.
\bibitem{27} J.C.Pati, A.Salam and J.Strathdee, Nuo.Cim. {\bf 26A}
        (1975) 72.
\bibitem{28} V.Elias, J.C.Pati and A.Salam, Phys.Rev.Lett.
        {\bf 40} (1978) 920.
\bibitem{29} R.Utiyama, Phys.Rev.{\bf 101} (1956) 1597.
\bibitem{30} A.Brodsky, D.Ivanenko and G. Sokolik, JETPH
        41 (1961)1307; Acta Phys.Hung. {\bf 14} (1962) 21.
\bibitem{31} T.W.Kibble, J.Math.Phys. {\bf 2} (1961) 212.
\bibitem{32} D.Ivanenko and G.Sardanashvily, Phys.Rep. {\bf 94}
        (1983) 1.
\bibitem{33} F.W.Hehl, J.D.McCrea, E.W.Mielke and Y.Ne'eman
    Phys.Rep. {\bf 258} (1995) 1-171.
\bibitem{34} F.W.Hehl, P. Von Der Heyde, G.D.Kerlick, J.M.Nester
    Rev.Mod.Phys. {\bf 48} (1976) 393-416.
\bibitem{35} Ning Wu, {\it Gauge Theory of Greavity},
        talk given
        at Meeting of the Devision of Particles and Fields
        of American Physical Society at the College of
        William \& Mery(DPF2002), May 24-28, 2002,
        Williamsburg, Virgia, USA; hep-th/0109145, hep-th/0207254.
\bibitem{36} Ning Wu, Commun. Theor. Phys. (Beijing, China)   
	{\bf 38} (2002): 151-156.
\bibitem{37} Ning Wu, Commun. Theor. Phys. (Beijing, China)  
	{\bf 38} (2002): 322-326.
\bibitem{38} Ning Wu, Commun. Theor. Phys. (Beijing, China)
	{\bf 38} (2002): 455-460
\bibitem{39} Ning Wu, {\it Quantization and Renormalization of
	$GSU(N)$ Unification Theory} (in preparation)
\bibitem{40} Ning Wu, (in preparation)









\end{thebibliography}
\end{document}